\documentclass[10pt]{bmc_article}

\usepackage{cite} 
\usepackage{url}  
\usepackage{ifthen}  
\usepackage{multicol}   
\usepackage[utf8]{inputenc} 
\usepackage{lineno,booktabs,graphicx}
\newboolean{publ}

\newenvironment{bmcformat}{\fussy\setboolean{publ}{true}}{\fussy}

\begin{document}
\begin{bmcformat}


\title{Improving basic and translational science by accounting for litter-to-litter variation in animal models}
 

\author{Stanley E. Lazic\correspondingauthor$^1$,\ %
       \email{Stanley E. Lazic\correspondingauthor - stan.lazic@cantab.net}%
       Laurent Essioux$^2$%
       \email{laurent.essioux@roche.com}%
      }


\address{%
    \iid(1) In Silico Lead Discovery, Novartis Institutes for Biomedical Research, Basel, Switzerland \\
    \iid(2) Bioinformatics and Exploratory Data Analysis, F. Hoffmann-La Roche, Basel, Switzerland
}%

\maketitle


\begin{abstract}
        \paragraph*{Background:}  Animals from the same litter are often more alike compared with animals from different litters. This litter-to-litter variation, or ``litter effects'', can influence the results in addition to the experimental factors of interest. Furthermore,  an experimental treatment can be applied to whole litters rather than to individual offspring. For example, in the valproic acid (VPA) model of autism, VPA is administered to pregnant females thereby inducing the disease phenotype in the offspring. With this type of experiment the sample size is the number of litters and not the total number of offspring. If such experiments are not appropriately designed and analysed, the results can be severely biased as well as extremely underpowered.
      
        \paragraph*{Results:}  A review of the VPA literature showed that only 9\% (3/34) of studies correctly determined that the experimental unit ($n$) was the litter and therefore made valid statistical inferences. In addition, litter effects accounted for up to 61\% (p $<$0.001) of the variation in behavioural outcomes, which was larger than the treatment effects. In addition, few studies reported using randomisation (12\%) or blinding (18\%), and none indicated that a sample size calculation or power analysis had been conducted.

        \paragraph*{Conclusions:}  Litter effects are common, large, and ignoring them can make replication of findings difficult and can contribute to the low rate of translating preclinical in vivo studies into successful therapies. Only a minority of studies reported using rigorous experimental methods, which is consistent with much of the preclinical in vivo literature.


\end{abstract}

\ifthenelse{\boolean{publ}}{\begin{multicols}{2}}{}


\section*{Background}

Numerous animal models (lesion, transgenic, knockout, selective breeding, etc.) have been developed for a variety of psychiatric, neurodegenerative, and neurodevelopmental disorders. While many of these models have been helpful for understanding disease pathology, they have been less useful for discovering potential therapies or predicting clinical efficacy. Translation from in vivo animal models (typically rodent) has been poor, despite many years of research and effort. There are many reasons for this, including the inherent difference in biology between rodents and humans \cite{Geerts2009}, particularly relating to higher cognitive functions. In addition, there is the ever-present question of whether a particular animal model is even suitable; whether it recapitulates the disease process of interest or faithfully mimics key aspects of the human condition. While important, these two considerations will be put aside and the focus will be on the design and analysis of preclinical studies using multiparous species, and how this affects the validity and reproducibility of results. There are two issues that will be discussed. The first deals with designs where an experimental treatment is applied to whole litters rather than to the individual animals, usually because the treatment is applied to pregnant females and therefore to all of the offspring. The second is the natural litter-to-litter variation that is often present, which means that the value of a measured experimental outcome can potentially be influenced by the litter that the animal came from.  \pb

\subsection*{Applying treatments to whole litters}
Some disease models have a distinctive experimental design feature: the treatment is applied to pregnant females (and therefore to all of the unborn animals within that female), but the scientific interest is in the individual offspring (Figure 1).  Here, the ``treatment'' refers to the experimental manipulation that induces the disease features, and it does not refer to a therapeutic treatment. This design is common in toxicology and nutrition studies, but is also used in neuroscience studies when examining the effects of maternal stress and in the valproic acid (VPA) model of autism. Difficulties arise because the experimental unit (``$n$''; defined as the smallest physical unit that can be randomly assigned to a treatment condition) is the pregnant dam and not the individual offspring \cite{Haseman1975,Hughes1979,Holson1992,Zorrilla1997,Wainwright1998,Festing2002,Festing2003,Festing2006,Casella2008,Maurissen2010,Lazic2010}. In other words, the sample size is the number of dams, and the offspring are considered subsamples, much like the left and right kidney from a single animal do not represent a sample size of two ($n$ = 1, even though there are two ``repeated'' measurements). This may come as a surprise, and it is irrelevant that the scientific interest is in the offspring, or that the offspring eventually become individual entities (unlike kidneys). Regulatory authorities have clear guidelines on the matter \cite{ICH1993,OECD2007}; for example, the Organisation for Economic Co-operation and Development (OECD) has made a firm statement in their guidelines for chemical testing: ``Developmental studies using multiparous species where multiple pups per litter are tested should include the litter in the statistical model to guard against an inflated Type I error rate. The statistical unit of measure should be the litter and not the pup. Experiments should be designed such that litter-mates are not treated as independent observations [p. 12]'' \cite{OECD2007}.  There is a restriction on randomisation because only whole litters can be assigned to the treatment or control conditions, which has implications for how studies are designed and analysed. An appropriate analysis can be conducted by using only one animal per litter (randomly selected), which allows standard methods to be used (e.g. t-test, ANOVA, etc.). This is often not the most efficient design in terms of animal usage, unless the excess animals can be used for other experiments. A second option is to use more than one animal per litter, and then average the values of the animals within a litter. These mean values can then be taken forward and analysed using standard methods. A third option is to use multiple animals per litter, and then use a mixed-effects model for analysis, which properly handles the structure of the data (i.e. animals are nested within litters) and avoids artificially inflating the sample size (also known as pseudoreplication \cite{Hurlbert1984,Lazic2010}). The third method is preferred to averaging values within a litter because the magnitude of the litter effect can be quantified. In addition, information on the precision of estimates will be lost by averaging, but is retained and made use of in the mixed-effects model. When using the first two options, it is clear that to increase the sample size and thus power, the number of litters needs to be increased. This is also true for the third option, but may not be so readily apparent \cite[pp. 3--4]{Casella2008}, and is discussed further below. \pb

A related design issue is that greater statistical power can be achieved when litter-mates are used to test a therapeutic compound versus a placebo. If the therapeutic treatment is applied to the individual animals postnatally, then the individual animal is the experimental unit \textit{for this comparison}. This is referred to as a split-plot design and has more than one type of experimental unit: litters for some comparisons and individual animals for others \cite{Casella2008}. These studies therefore require careful planning and analysis, but biologists are rarely introduced to these designs and how to appropriately analyse data derived from them during the course of their training. \pb

\subsection*{Litter effects are ubiquitous, large, and important}
It is known that for many measurable characteristics across many species, monozygotic twins are more similar than dizygotic twins, which are more similar than non-twin siblings, and which in turn are more similar than two unrelated individuals. What has not been fully appreciated is that all of the standard statistical methods (e.g. t-test, ANOVA, regression, non-parametric methods) assume that the data come from unrelated individuals. However, rodents from the same litter are effectively dizygotic twins; they are genetically very similar and share prenatal and early postnatal environments. Therefore, studies need to be designed and analysed in such a way that if differences between litters exist, they do not bias or confound the results \cite{Haseman1975,Hughes1979,Holson1992,Zorrilla1997,Wainwright1998,Festing2002,Festing2003,Festing2006,Casella2008,Maurissen2010,Lazic2010}. More specifically, this relates to the assumption of independence of observations. For example, measuring blood pressure (BP) from the left and right arm of ten unrelated people only provides ten independent measurements of BP, not twenty. This is because the left and right BP values will be highly correlated---if the BP value measured from a person's left arm is high, then so will the value measured from their right arm. Similarly, two animals from the same litter will tend to have values that are more alike (i.e. correlated) than two animals from different litters. One can think of it as within-litter homogeneity and between-litter heterogeneity. This lack of independence needs to be handled appropriately in the analysis and the three strategies outlined in the previous section can be used. Many animal models are derived from highly inbred strains, and this results in reduced genotypic and phenotypic variation. This is a different issue and unrelated to lack of independence. It does not mean that animals ``are all the same'' and that differences between litters do not exist. \pb

Litter effects are not a minor issue that only statistical pedants worry about with little practical importance for scientists. Using actual body weight data from their experiment, Holson and Pearce showed that if three ``treated'' and three ``control'' litters are used, with two offspring per litter (total number of offspring = 12), then the false positive rate (Type I error) is 20\% rather than the assumed 5\% \cite{Holson1992}. The false positive rate was determined as the proportion of p-values that fell below 0.05. Since there was no actual treatment applied, random sampling should produce only a 5\% error rate. Furthermore, they showed that the false positive rate increases with the number of offspring per litter; if the number of offspring per litter is 12 (total number of offspring = 72) then the false positive rate is 80\%. The error rate is also influenced by the relative variability between and within litters and will therefore vary for each experimental outcome. Given that papers report the results of multiple tests (multiple outcome variables and multiple comparisons), we can expect the literature to contain many false positive results. It may seem paradoxical, but in addition to an increased false positive rate, ignoring litter-to-litter variation can also lead to low power (too many false negatives) when true effects exist \cite{Holson1992,Zorrilla1997}. This occurs because litter-to-litter variation is unexplained variation, and thus the ``noise'' in the data is increased, potentially masking true treatment effects. A subsequent study using forty litters found ``significant litter effects\dots\ in varying degrees, for almost every behavioural, morphologic, and neuroendocrine measure; they were evident across indices of neural, adrenal, thyroid, and immunologic functioning in adulthood'' \cite{Zorrilla1997} (and see references therein for further studies supporting this conclusion). Holson and Pearce reported that only 30\% of papers in the behavioural neurotoxicology literature correctly accounted for litter effects \cite{Holson1992} and Zorrilla  noted that 34\% of papers in \textit{Developmental Psychobiology} and 15\% of papers in related journals correctly accounted for litter effects \cite{Zorrilla1997}. This issue has been discussed repeatedly for almost forty years \cite{Haseman1975}, but experimental biologists seem unaware (or choose to ignore) the importance of dealing with litter effects. One can only speculate on the number of erroneous conclusions that have been reached and the resources that have been wasted. \pb

One might argue that when many studies are conducted, including replications within and between labs, the evidence will eventually converge to the ``truth'', and therefore these considerations are only of minor interest. Unfortunately, there is no guarantee of such convergence, as the literature on the superoxide dismutase (SOD1) transgenic mouse model of amyotrophic lateral sclerosis (ALS) demonstrates.  Several treatments showed efficacy in this model and were advanced to clinical trials, where they proved to be ineffective \cite{Schnabel2008}. A subsequent large-scale and properly executed replication study did not support the previous findings \cite{Scott2008}. This study also identified litter as an important variable that affected survival (the main outcome) and which was not taken into account in the earlier studies. The authors also demonstrated how false positive results can arise with inappropriate experimental designs and analyses. Litter effects were not the only contributing factor; a meta-analysis of the preclinical SOD1 literature revealed that only 31\% of studies reported randomly assigning animals to treatment conditions and even fewer reported blind assessment of outcomes \cite{Benatar2007}.  Lack of randomisation and blinding are known to overstate the size of treatment effects \cite{Bebarta2003,Crossley2008,Kilkenny2009,Sena2010,Vesterinen2010}. In addition, there was evidence of publication bias, where studies with positive results were more likely to be published \cite{Benatar2007}. Thus, the combination of poor experimental design, analysis, and publication bias contributed to numerous incorrect decisions regarding treatment efficacy. \pb

\subsection*{General quality of preclinical animal studies}
Previous studies have shown that the general quality of the design, analysis, and interpretation of preclinical animal experiments is low \cite{Hackam2006,Dirnagl2006,Crossley2008,Kilkenny2009,Philip2009,Vesterinen2010,Worp2010,Shineman2011,Rooke2011,Sarewitz2012}. For example, Nieuwenhuis et al. recently reported that 50\% of papers in the neuroscience literature misinterpret interaction effects \cite{Nieuwenhuis2011}. In addition, the issue of ``inflated $n$'', or pseudoreplication, shows up in other guises \cite{Cumming2007,Lazic2010}, and whole fields can misattribute cause-and-effect relationships \cite{Lazic2010a,Lazic2011}. There is also the concept of ``researcher degrees of freedom'', which refers to the post hoc flexibility in choosing the main outcome variables, statistical models, data transformations, how outliers are handled, when to stop collecting data, and what is reported in the final paper \cite{Simmons2011}. Various permutations of the above options greatly increases the chances that something statistically significant can be found, and this gets reported as the sole analysis that was conducted. Given the above concerns, it is not surprising that the pharmaceutical industry has difficulty reproducing many published results \cite{Mullard2011,Prinz2011,Ledford2012,Begley2012}. \pb

\section*{Methods}

\subsection*{Literature review}
Ninety-five papers were identified on PubMed using the search term ``(VPA [tiab] OR `valproic acid' [tiab]) AND autism [tiab]'' (up to the end of 2011). Reference lists from these articles were then examined for further relevant studies, and one was found. Only primary research articles that injected pregnant dams with VPA and subsequently analysed the effects in the offspring were selected (in vitro studies were excluded). A total of thirty-five studies were found, and one was excluded as key information was located in the supplementary material, but this was not available online \cite{Rinaldi2008}. Two key pieces of information were extracted: (1) whether the analysis correctly identified the experimental unit as the litter and (2) whether important features of good experimental design were mentioned, including randomisation, blinding, sample size calculation, and whether the total sample size (i.e. number of pregnant dams) was indicated or could be determined. \pb

\subsection*{Estimating the importance of litter-to-litter variation}
Data from Mehta et al. \cite{Mehta2011} were used to estimate the magnitude of differences between litters on a number of outcome variables. This study was chosen because it included animals from fourteen litters (five saline, nine VPA) and therefore it was possible to get a reasonable estimate of the litter-to-litter variation. In addition, the study mentioned using randomisation and blind assessment of outcomes.  Half of the animals in each condition were also given MPEP (2-methyl-6-phenylethyl-pyrididine), a metabotropic glutamate receptor 5 antagonist. To assess the magnitude of the litter effects, the effect of VPA,  MPEP, and sex (if relevant) were removed, and the remaining variability in the data that could be attributed to differences between litters was estimated. More specifically, models with and without a random effect of litter were compared with a likelihood ratio test. This analysis is testing whether the variance between litters is zero, and it is known that p-values will be too large because of ``testing on the boundary'', and so the simple method of dividing the resulting p-values by two was used as recommended by Zuur et al \cite{Zuur2009}. The exact specification of the models is provided as R code in Additional File 1 and the data are provided in Additional File 2. \pb

\subsection*{Power analysis}
In these types of designs, power (the ability to detect an effect that is actually present) is influenced by (1) the number of litters, (2) variability between litters, (3) number of animals within litters, (4) variability of animals within litters, (5) difference between the means of the treatment groups (effect size), (6) significance cutoff (traditionally $\alpha$ = 0.05), and (7) the statistical test used. In order to illustrate the importance of the number of litters relative to the number of animals within litters and how an inappropriate analysis can lead to p-values that are too small, a power analysis was conducted with the number of litters per group varying from three to ten, and the number of animals per litter varying from one to ten. The other factors were held constant. Variability between litters (SD = 0.8803) and the variability of animals within litters (SD = 0.8142) was estimated from the locomotor activity data from Mehta et al. \cite{Mehta2011}.  For each combination of litters and animals, 5000 simulated datasets were created with a mean difference between groups of 0.15. Once the datasets were generated, the power for three types of analyses were calculated. The first analysis averaged the values of the animals within each litter and then groups were compared with a t-test. The second analysis used a mixed-effects model, and the third ignored litter and compared all of the values with a t-test. The last analysis is incorrect and only presented to demonstrate how artificially inflating sample size affects power. The power for each analysis was determined as the proportion of tests that had p $<$ 0.05. The R code is provided in Additional File 1 and is adapted from Gelman and Hill \cite{Gelman2007}. \pb

\section*{Results and Discussion}
\subsection*{Low quality of the published literature}
The VPA model of autism is relatively new and potential therapeutic compounds tested in this model have not yet advanced to human trials. The opportunity therefore exists to clean up the literature and prevent a repeat of the SOD1 story. The main finding is that only 9\% (3/34) of studies correctly identified the experimental unit and thus made valid inferences from the data. One study used a nested design \cite{Stodgell2006}, the second mentioned that litter was the experimental unit \cite{Kuwagata2009}, and the third used one animal from each litter, thus bypassing the issue \cite{Murawski2009}. In fourteen studies (41\%) it was not possible to determine the number of dams that were used (i.e. the sample size) and in four studies (12\%) the number of offspring used were not indicated. In addition, only four (12\%) reported randomly assigning pregnant females to the VPA or control group. Many studies also used only a subset of the offspring from each litter, but often it was not mentioned how the offspring were selected. Only six studies (18\%) reported that the investigator was blind to the experimental condition when collecting the data. Ten studies (29\%) did not indicate whether both male and female offspring were used. No study mentioned performing a power analysis to determine a suitable sample size to detect effects of a given magnitude---but this is probably fortuitous, given that only three studies correctly identified the experimental unit. It is possible that many studies did use randomisation and assess outcomes blindly, but simply did not report it. However, randomisation and blinding are crucial aspects for the validity of the results and their omission in  manuscripts suggests that they were not used. This is further supported by studies showing that when manuscripts do not mention using randomisation or blinding the estimated effects sizes are larger compared to studies that do mention using these methods, which is suggestive of bias \cite{Bebarta2003,Crossley2008,Kilkenny2009,Sena2010,Vesterinen2010,Rooke2011}. \pb

A number of papers had additional statistical or experimental design issues, ranging from trivial (e.g. reporting total degrees of freedom rather than residual degrees of freedom for an F-statistic) to serious. These include treating individual neurons as the experimental unit, which is common in electrophysiological studies, but just as inappropriate as treating blood pressure values taken from left and right arms as $n=2$, or dissecting a single liver sample into ten pieces and treating the expression of a gene measured in each piece as $n=10$ \cite{Lazic2010}. If it were that easy, clinical trials could be conducted with tens of patients rather than hundreds or thousands.  Regulatory authorities are not fooled by such stratagems, but is seems many journal editors and peer-reviewers are.  A list of studies can be found in Additional File 3. \pb

\subsection*{Estimating the magnitude of litter effects}
To illustrate the extent to which litter effects can influence the results, data originally published by Mehta et al. \cite{Mehta2011} were used and experimental details can be found therein. Locomotor activity in the open field is shown in Figure 2 for nine VPA and five saline injected control litters. Half of the animals from each condition were given MPEP (a mGluR5 receptor antagonist) or saline. Visually, there do not appear to be differences between VPA and control groups and there is a slight increase in activity due to MPEP. The effect of MPEP was not significant when litter effects were ignored (Figure 2A; p = 0.082), but it was when adjusting for litter (Figure 2B; p = 0.011). In this case the shift in p-value was not large, but it happened to decrease it below the 0.05 threshold after the excess noise caused by litter-to-litter variation was removed.  \pb

It may be difficult to determine whether litter effects are present by simply plotting the data by litter because they may be obscured by the experimental effects. For a visual check, it is preferable to remove the effect of the experimental factors first and then plot the residual values versus litter. The $y$-axis for Figure 3 shows the residuals, which are defined as the difference between the observed locomotor activity for each animal and the value predicted from the model containing group (VPA/saline) and condition (MPEP/saline) as factors (from Figure 2A). The residuals should be pure noise, centred at zero, and should not be associated with any other variable. However, it is clear that there are large differences between litters (Figure 3A), indicating heterogeneity in the response from one litter to the next. When litter effects are taken into account, the mean of each litter is closer to zero. Also note that variance of the residuals ($\sigma^2_\epsilon$) is reduced by 61\% when litter is taken into account (p $<$ 0.001). This is shown by the spread of the grey points around zero on the right side of each graph, which are clustered closer together in the second analysis. The interpretation is that litter accounted for 61\% of the previously unexplained variation in the data. Note that it would be impossible to determine whether litter effects are present if only one litter per treatment group was used because litter and treatment would be completely confounded.  \pb

A similar analysis was performed for other variables and the results are displayed in Table 1. It is clear that litter-to-litter variation is important for a number of behavioural outcomes. It is also clear from Figure 3A how one could obtain false positives with an inappropriate design and analysis. Suppose an experiment was conducted with only one VPA and one saline litter, with ten animals from each, and that there is no overall effect of VPA on a particular outcome. If the experimenter happened to select Litter A (saline) and Litter M (VPA) there would be a significant increase due to VPA, but if Litter D (saline) and Litter G (VPA) were selected, there would be a significant effect in the opposite direction! There are many combinations of a single saline and VPA litter that would lead to a significant difference between conditions. Having two or three litters per group instead of one will reduce the false positive rate, but it will still be much higher than 0.05 \cite{Holson1992}. In addition, these apparent differences would not replicate with a properly designed follow-up experiment.  \pb

\subsection*{How power is affected by the number of litters and animals}
Figure 4 shows the power for various combinations of number of litters and number of animals per litter. This analysis is based on averaging the values for the animals within a litter and then comparing the groups with a t-test. It is clear that increasing the number of animals per litter has little effect on power (the lines in Figure 4A are nearly flat after two animals per litter), whereas increasing the number of litters results in a large increase in power. The results for the mixed-effect model are nearly identical and the results of the inappropriate analysis which ignores litter shows increasing power with increasing number of animals per litter (Additional File 4). This is false power however, and is due to an artificially inflated sample size (pseudoreplication) that will lead to many false positive results.\pb

Some may object on ethical grounds to using so many litters and then selecting only one or a few animals from each, as there will be many additional animals that will not be used and presumably culled. Certainly all of the animals could be used, but there is almost no increase in power after three animals per litter (at least for the locomotor data) and therefore it is a poor use of time and resources to include all of the animals. One could argue therefore that it is unethical to submit a greater number of animals to the experimental procedure if they contribute little or nothing to the result. One could also argue that it is even more unethical to use any animals for a severely underpowered (or flawed) study in the first place and then to clutter the scientific literature with the results. One way to deal with the excess animals is to use them for other experiments. This requires greater planning, organisation, and coordination, but it is possible. Another option is to purchase animals from a commercial supplier and request that the animals come from different litters rather than have an in-house colony. As a side note, suppliers do not routinely provide information on the litters that the animals came from and thus an important variable is not under the experimenter's control and cannot even be checked whether it is influencing the results.  \pb

\subsection*{How does litter-to-litter variation arise?}
Differences between litters could exist for a variety of reasons, including shared genes and shared prenatal and early postnatal environments, but also due to age differences (it is difficult to control the time of mating), and because litters are convenient units to work with. For example, it is not unusual for litter-mates to be housed in the same cage, which means that animals within a litter also share not just their early, but also their adult environment. It is also often administratively easier to apply experimental treatments on a per cage (and thus per litter) basis rather than per animal basis. For example, animals in cage A and C are treated while cage B and D are controls. Animals may also undergo behavioural testing on a per cage basis; for example, animals are taken from the housing room to the testing room one cage at a time, tested, and then returned. Larger experiments may need to be conducted over several days and it is often easier to test all the animals in a subset of cages on each day, rather than a subset of animals from all of the cages. At the end of the experiment animals may also be killed on a per cage basis. Given that it may take many hours to kill the animals, remove the brains, collect blood, etc., the values of many outcomes such as gene expression, hormone and metabolite concentrations, and physiological parameters may change due to circadian rhythms. All of these can lead to systematic differences between litters and can thus bias results and/or add noise to the data.\pb

There is an important distinction to be made between applying treatments to whole litters versus ``natural'' variation between litters. When a treatment is applied to a whole litter such as the VPA model of autism or maternal stress models, then the litter is the experimental unit and the sample size is the number of litters. Therefore, \emph{by definition}, litter needs to be included in the analysis if more than one animal per litter is used (or the values within a litter can be averaged).  However, if multiple litters are used but the treatment(s) are applied to the individual animals, experiments should be designed so that \emph{if} litter effects exist, then valid inferences can still be made. In other words, litters should not be confounded with other experimental variables because it would be difficult or impossible to detect their influence and remove their effects. Whether litter is an important factor for any particular outcome is then an empirical question, and if it is not important then it need not be included in the analysis. However, the power to detect differences between litters will be low if only a few litters are used in the experiment and therefore a non-significant test for litter effects should not be interpreted as the absence of such effects. Analysing the data with and without litter and choosing the analysis that gives the ``right'' answer should of course be avoided \cite{Simmons2011}. Flood et al. provide a nice example in the autism literature of an appropriate design followed by a check for litter effects, and then the results for the experimental effect were reported when litter was both included and excluded \cite{Flood2012}. Consistent with other studies demonstrating litter-effects, this paper found a strong effect of litter on brain mass. \pb

\subsection*{Four ways to improve basic and translational research}
\subsubsection*{Better training for biologists}
Most experimental biologists are not provided with sufficient training in experimental design and data analysis to be able to plan, conduct, and interpret the results of scientific investigations at the level required to consistently obtain valid results. The solution is straightforward, but requires major changes in the education and training of biologists and it will take many years to implement. Nevertheless, this should be a long-term goal for the biomedical research community. \pb

\subsubsection*{Make better use of statistical expertise}
A second solution is to have statisticians play a greater role in preclinical studies, including peer reviewing grant applications and manuscripts, as well as being part of scientific teams \cite{Peers2012}. However, there are not enough statisticians with the appropriate subject matter knowledge to fully meet this demand---just as it is difficult to do good science without a knowledge of statistics, it is difficult to perform a good analysis without knowledge of the science. In addition, this type of ``project support'' is often viewed by academic statisticians as a secondary activity. Despite this, there is still scope for improving the quality of studies by making better use of statistical expertise. \pb

\subsubsection*{More detailed reporting of experimental methods}
Detailed reporting of how experiments were conducted, how data were analysed, how outliers were handled, whether all animals that entered the study completed it, and how the sample size was determined are all required to assess whether the results of the study are valid, and a number of guidelines have been proposed which cover these points, including the National Institute of Neurological Disorders and Stroke (NINDS) guidelines \cite{Landis2012}, the Gold Standard Publication Checklist \cite{Hooijmans2010}, and the ARRIVE (Animals in Research: Reporting In Vivo Experiments) guidelines \cite{Kilkenny2010}. For example, ARRIVE items 6 (Study design), 10 (Sample size), 11 (Allocating animals to experimental groups), and 13 (Statistical methods) should a be mandatory requirement for all publications involving animals and could be included as a separate checklist that is submitted along with the manuscript, much like a conflict of interest or a transfer of copyright form. Something similar has recently been introduced by \textit{Nature Neuroscience} \cite{NN2013}. This would make it easier to spot any design and analysis issues by reviewers, editors, and other readers. In addition, and more importantly, if scientists are required to comment on how they randomised treatment allocation, or how they ensured that assessment of outcomes was blinded, then they will conduct their experiments accordingly if they plan on submitting to a journal with these reporting requirements. Similarly, if researchers are required to state what the experimental unit is (e.g. litter, cage, individual animal, etc.), then they will be prompted to think hard about the issue and design better experiments, or seek advice. This recommendation will not only improve the quality of reporting, but it will also improve the quality of experiments, which is the real benefit.  A final advantage is that it will make quantitative reviews/meta-analyses easier because much of the key information will be on a single page. \pb

\subsubsection*{Make raw data available}
Another solution is to make the provision of raw data a requirement for acceptance of a manuscript; not ``to make it available if someone asks for it'', which is the current requirement for many journals, but uploaded as supplementary material or hosted by a third party data repository. None of the VPA studies provided the data that the conclusions were based on, making reanalysis impossible. Remarkably, of the thirty-five studies published, only one provided the necessary information to conduct a power analysis to plan a future study\cite{Murawski2009}, and this was only because one animal per litter was used and the necessary values could be extracted from the figures. Datasets used in preclinical animal studies are typically small, do not have confidentiality issues associated with them, are unlikely to be used for further analyses by the original authors, and have no additional intellectual property issues associated with them given that the manuscript itself has been published. It is noteworthy that many journals require microarray data to be uploaded to a publicly available repository (e.g. Gene Expression Omnibus or ArrayExpress), but not the corresponding behavioural or histological data. It is perhaps not surprising that there is a relationship between study quality and the willingness to share data \cite{Wicherts2006,Bakker2011,Wicherts2011}. Publishing raw data can be taken as a signal that researchers stand behind their data and therefore their conclusions. Funding bodies should encourage this by requiring that data arising from the grant are made publicly available (with penalties for non-adherence). \pb

The above suggestions would help ensure that appropriate design and analyses were used, and to make it easy to verify claims or to reanalyse data. Currently, it is often difficult to establish the former and almost impossible to perform the latter. Moreover, it is clear that appropriate designs and analyses are often not used, making it difficult to give the benefit of the doubt to those studies with incomplete reporting of how experiments were conducted and data analysed. \pb

\section*{Conclusions}
While it is difficult to quantify the extent to which poor statistical practices hinder basic and translational research, it is clear that a large inflation of false positive and false negative rates will only slow progress. In addition, because of publication bias and researcher degrees of freedom, it is possible for a field to converge to the wrong answer. Experimental design and statistical issues are, in principle, fixable. Improving these will allow scientists to focus on creating and assessing the suitability of disease models and the efficacy of therapeutic interventions, which is challenging enough. \pb

\section*{List of abbreviations}
ANOVA: analysis of variance; BP: blood pressure; MPEP: 2-methyl-6-phenylethyl-pyrididine; SOD1: superoxide dismutase; VPA: valproic acid

\section*{Authors contributions}
SEL planned and carried out the study, performed the literature search and analysis, and wrote the paper. LE provided constructive input. All authors read and approved the final manuscript.

\section*{Competing interests}
The author declares no competing interests.

\section*{Acknowledgements}
 \ifthenelse{\boolean{publ}}{\small}{}
The authors would like to thank the Siegel lab at the University of Pennsylvania for kindly sharing their data.


{\ifthenelse{\boolean{publ}}{\footnotesize}{\small}
 \bibliographystyle{bmc_article}  
  \bibliography{VPA} }     


\ifthenelse{\boolean{publ}}{\end{multicols}}{}


\clearpage
\section*{Figures}
  \subsection*{Figure 1 - Defining the experimental unit}
Pregnant females are the experimental units because they are randomised to the treatment (e.g. valproic acid) or control conditions and therefore $n = 6$ in this example. The three offspring within a litter will often be more alike than offspring from different litters $\left( \frac{\mathrm{Between~litter\ variation}}{\mathrm{Within~litter\ variation}} > 1\right)$ and multiple offspring within a litter can be thought of as subsamples or ``technical replicates'', even though these are the scientific unit of interest. Only the mean of the within-litter values are important when comparing treated and control groups. Using all of the offspring without averaging will result in an inflated sample size (pseudoreplication) with standard statistical analyses. Instead of averaging, one could randomly select only one animal from each litter, or use a mixed-effects model to appropriately partition the different sources of variation. The only way to increase sample size, and thus power, is to increase the number of litters used.

\begin{figure}[ht]
\centering
\includegraphics[scale=0.65]{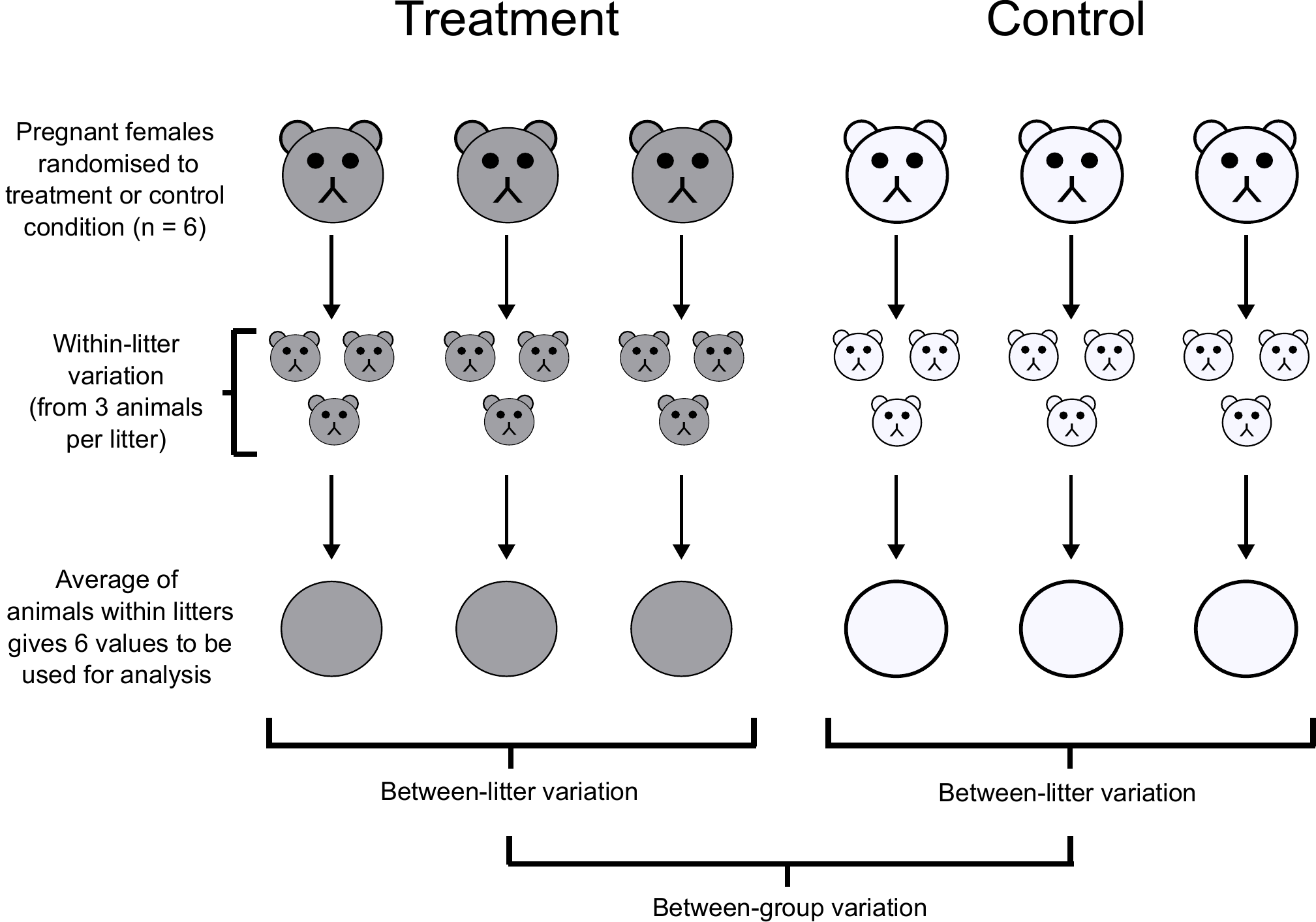}
\end{figure}

\clearpage
  \subsection*{Figure 2 - Analysis with and without litter taken into account} Nine pregnant female C57BL/6 mice were injected with 600 mg/kg VPA subcutaneously on embryonic day 13, and five control females received vehicle injections. Half of the animals in each condition were also injected with either a mGluR5 receptor antagonist (MPEP) or saline postnatally. Total locomotor activity in the open field over a 30\,min period at 8--9 weeks of age is shown. There was a slight increase in activity due to MPEP, but it was not significant when differences between litters were ignored (A; two-way ANOVA, mean difference = 0.60, F(1,44) = 3.17, p = 0.082).  Adjusting for litter removed unexplained variation in the data, allowing the small difference between groups to become statistically significant (B; mixed-effects model, mean difference = 0.64,  F(1,32) = 7.19, p = 0.011). Note how the values in the second graph have less variability around the group means; this increased precision leads to greater power of the statistical tests. Lines go through the mean of each group and points are jittered in the $x$ direction.

\begin{figure}[ht]
\centering
\includegraphics[scale=0.65]{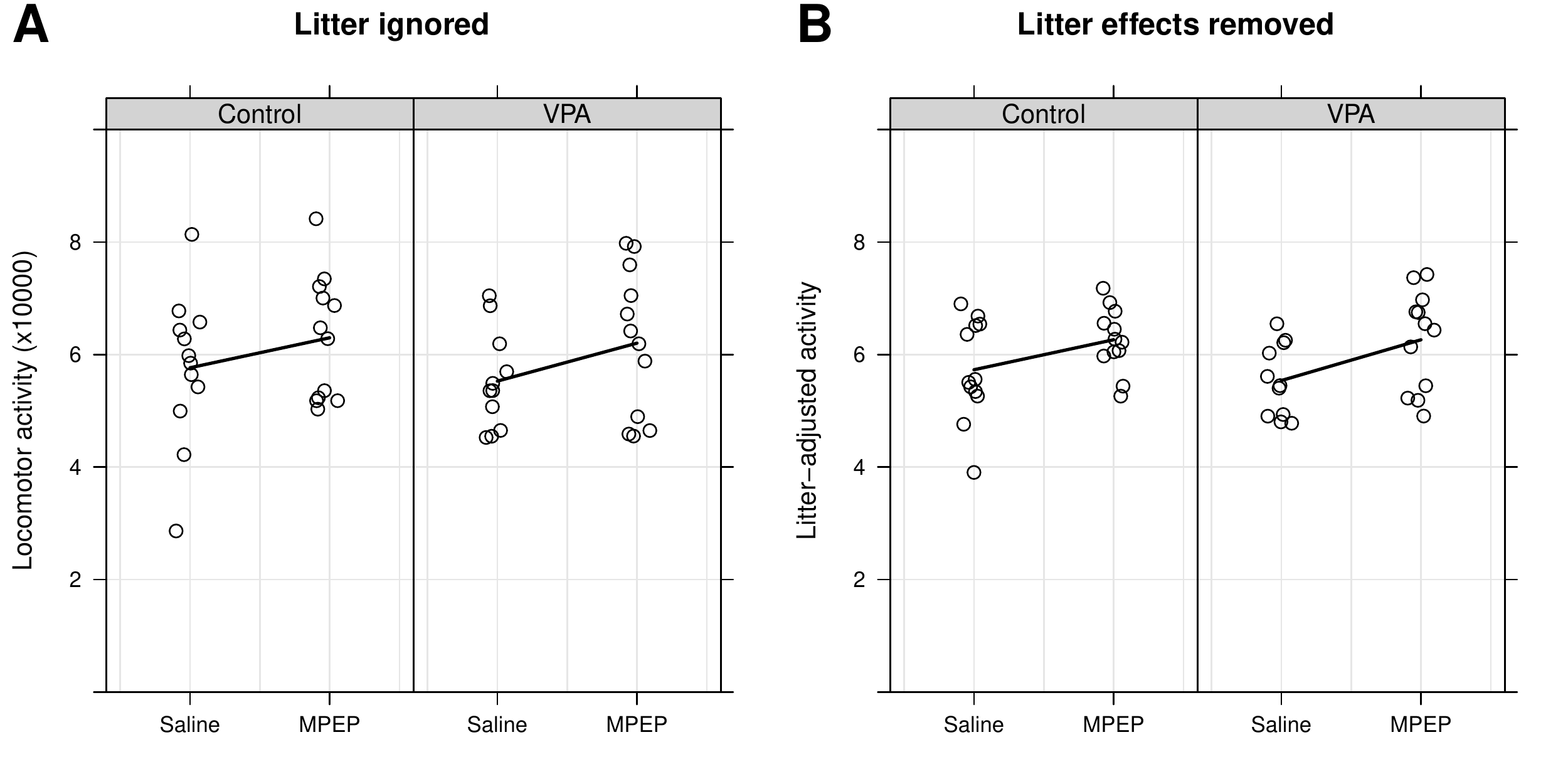}
\end{figure}

\clearpage
  \subsection*{Figure 3 - Visualising litter-to-litter variation} The residuals represent the unexplained variation in the data after the effects of VPA and MPEP have been taken into account; they should be pure noise and therefore not associated with any other variable. However, the standard analysis (A) shows that when residuals are plotted against litter ($x$-axis) there are large differences between litters. In other words, there is another factor affecting the outcome besides the experimental factors of interest. The variance of the residuals (grey points on the right) is high ($\sigma^2_\epsilon$ = 1.29). The proper analysis (B) reduces the unexplained variation in the data by 61\% ($\sigma^2_\epsilon$ = 0.50; p $<$ 0.001), which can be seen by the narrower spread of the grey points around zero, and the large differences between the litters have been removed. This reduction in noise allows smaller true signals to be detected. Error bars are SEM. Litters F and L only have one observation and thus no error bars.

\begin{figure}[ht]
\centering
\includegraphics[scale=0.55]{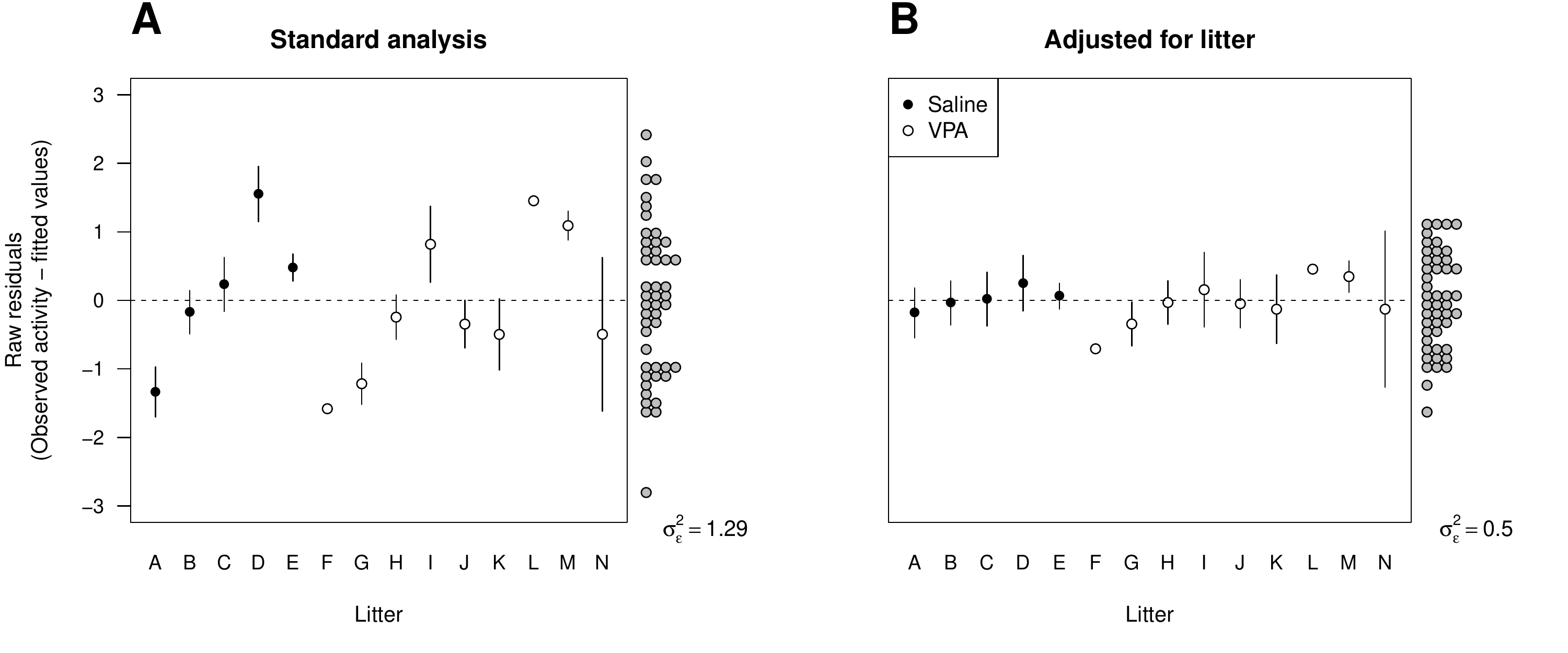}
\end{figure}

\clearpage
\subsection*{Figure 4 - Power calculations for VPA experiments} 
Panel A shows how power changes as the number of animals per litter increases from one to eight ($x$-axis) and the number of litters per group increases from three to ten (different lines). It is clear that increasing the number of animals per litter has only a modest effect on power with little improvement after two animals. A two-group study with three litters per group and eight animals per litter ($2 \times 3 \times 8 = 48$ animals) will have only a 30\% chance of detecting the effect, whereas a study with ten litters per group and one animal per litter ($2 \times 10 \times 1 = 20$ animals) will have almost 80\% power and also use far fewer animals.  Panel B shows the same data, but presented differently. Power for different combinations of litters and animals per litter is indicated by colour (red = low power, white = high) and reference lines for 70\%, 80\%, and 90\% power are indicated. Note that these specific power values are only relevant for the locomotor activity task with a fixed effect size and will have to be recalculated for other outcomes. However, the general result (increasing litters is better than increasing the number of animals per litter) will apply for all outcomes.

\begin{figure}[ht]
\centering
\includegraphics[scale=0.6]{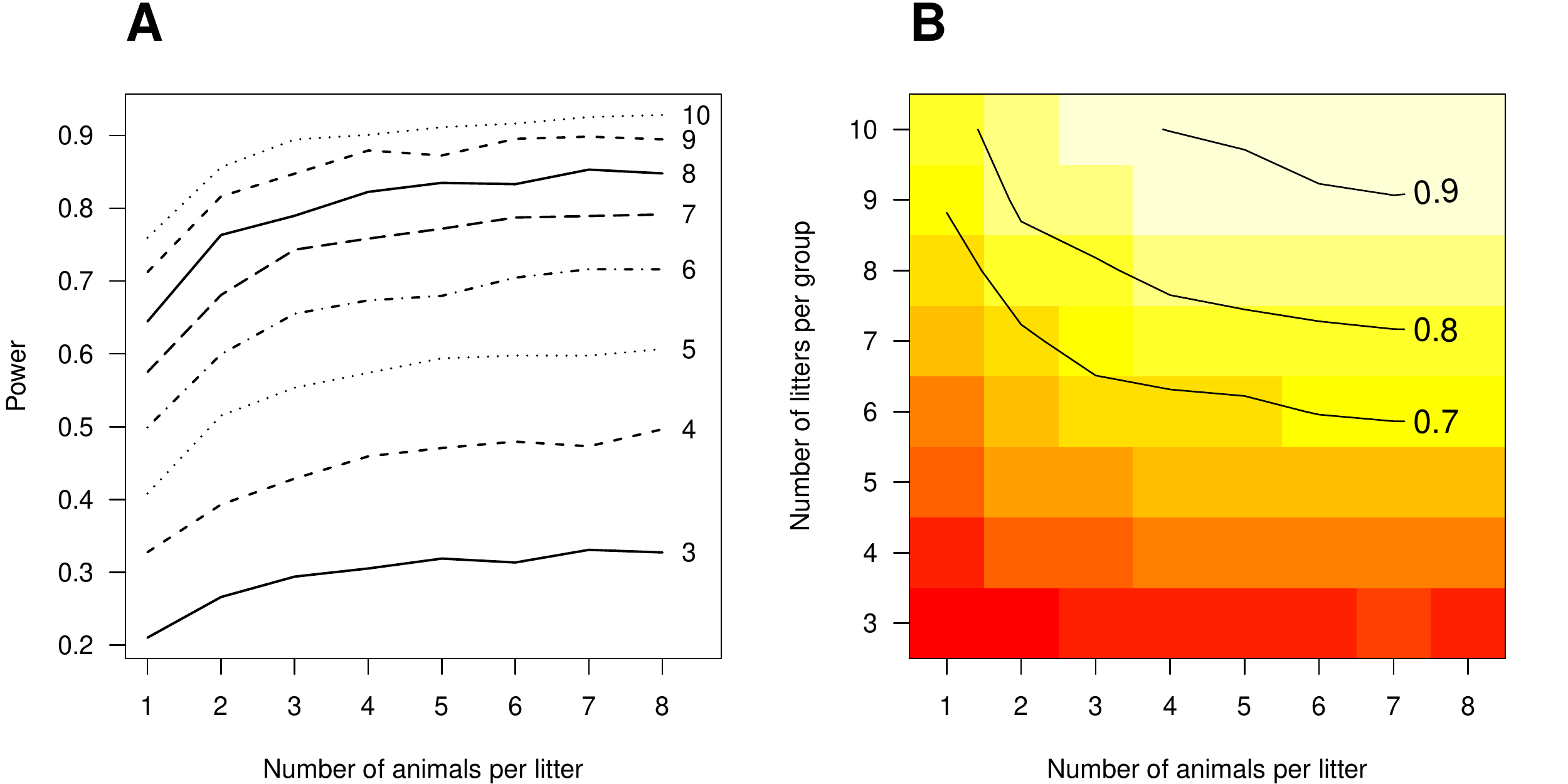}
\end{figure}



\clearpage
\section*{Tables}

\subsection*{Table 1 - Importance of litter effects on body weight and behavioural tests.}
The p-value tests whether the litter-to-litter variation was significantly greater than zero.

\begin{center}
\par
\mbox{ 
  \begin{tabular}[c]{lrr} \toprule
    \textbf{Variable} &  \textbf{Reduction in $\sigma^2_\epsilon$} & \textbf{P-value} \\ \midrule
    Locomotor activity & 61\% &  $<$0.001 \\
    Body weight  & 50\% & 0.003 \\
    Marbles buried & 38\% & 0.045 \\  
    Anxiety (open field)& 35\% & 0.0504 \\  
    Grooming & 23\% & 0.116 \\ \bottomrule
  \end{tabular} 
}\\
 $\sigma^2_\epsilon$ is the residual (unexplained) variation.
\end{center}



\section*{Additional Files}
  \subsection*{Additional file 1 --- R code for the analyses and power calculations}
Code for the analyses and power calculations are given as a plain text file.

  \subsection*{Additional file 2 --- Raw data}
Raw data from Mehta et al. \cite{Mehta2011}, including body weight, locomotor activity and anxiety measures from the open field test, grooming behaviour, and number of marbles buried in the marble-burying test. Details can be found in the original publication.

  \subsection*{Additional file 3 --- List of VPA studies}
List of the thirty-four studies using the VPA rodent model of autism.

  \subsection*{Additional file 4 --- Power analysis for the mixed-effects model and the incorrect analysis}
The interpretation of the graphs is the same as Figure 4 (main text). Panels A and B are for the mixed-effects model and are nearly identical to the results for averaging the values within each litter and then using a t-test (Figure 4 main text). Panels C and D ignore litter and compare all of the data with a t-test, which results in an artificially inflated sample size and inappropriately high power.

\end{bmcformat}
\end{document}